\documentclass[conference]{IEEEtran}

\usepackage{color}
\usepackage{cite}
\usepackage{fixltx2e}
\usepackage[cmex10]{amsmath}
\usepackage{array}
\usepackage{wasysym}
\usepackage{dsfont}
\usepackage[latin1]{inputenc}                       
\usepackage[english]{babel}                         
\usepackage[T1]{fontenc}
\usepackage{mathtools}
\usepackage{amssymb} 
\usepackage{enumerate}
\usepackage{bbm}
\usepackage{epsfig,syntonly}
\usepackage{verbatim,times}
\usepackage{epstopdf}
\usepackage{graphicx}
\usepackage{latexsym,fancyhdr,bm}
\usepackage[tight,footnotesize]{subfigure}
\usepackage{url}
\usepackage{bbm}

\newtheorem{definition}{Definition}

\newtheorem{theorem}{Theorem}

\newtheorem{lemma}{Lemma}

\newcommand{\code}{{\mathcal C}}

\newcommand{\prob}{{\mathbb P}}

\newcommand{\cw}{c}

\makeatletter
\let\l@ENGLISH\l@english
\makeatother

\title{Reed-Muller Codes Achieve Capacity on the Binary Erasure
Channel under MAP Decoding}
\author{\IEEEauthorblockN{Shrinivas Kudekar\IEEEauthorrefmark{3}, Marco Mondelli\IEEEauthorrefmark{1}, Eren \c{S}a\c{s}o\u{g}lu\IEEEauthorrefmark{2}, R\"{u}diger Urbanke\IEEEauthorrefmark{1}, }
\IEEEauthorblockA{\IEEEauthorrefmark{1}School of Computer and Communication Sciences, EPFL, Switzerland\\
Emails: \texttt{\{marco.mondelli, ruediger.urbanke\}@epfl.ch}}
\IEEEauthorblockA{\IEEEauthorrefmark{2}UC Berkeley\\
Email: \texttt{eren.sasoglu@gmail.com}}
\IEEEauthorblockA{\IEEEauthorrefmark{3}Qualcomm Research, New Jersey, USA\\
Email: \texttt{skudekar@qti.qualcomm.com}}
}

\begin{document}

\maketitle
%------------------------------------------------------------------------------------------------------------------
\begin{abstract}
\noindent 
We show that Reed-Muller codes achieve capacity under maximum a
posteriori bit decoding for transmission over the binary erasure
channel for all rates $0 < R < 1$. The proof is generic and applies
to other codes with sufficient amount of symmetry as well. The main
idea is to combine the following observations: (i) monotone functions
experience a sharp threshold behavior, (ii) the extrinsic information
transfer (EXIT) functions are monotone, (iii) Reed--Muller codes
are 2-transitive and thus the EXIT functions associated with their
codeword bits are all equal, and (iv) therefore the Area Theorem
for the average EXIT functions implies that RM codes' threshold is
at channel capacity.  \end{abstract}

\begin{IEEEkeywords}
RM codes, MAP decoding, capacity-achieving codes, BEC, EXIT function
\end{IEEEkeywords}

%----------------------------------------------------
\section{Introduction} \label{sec:intro}
%----------------------------------------------------

Reed--Muller (RM) codes \cite{Mul54,Ree54,sloane,dumer:RMalgo} are among the
oldest codes in existence, and due to their many desirable properties,
are also among the most widely studied. In recent years there has
been renewed interest in RM codes, partly due to the invention of
capacity-achieving polar codes~\cite{Ari09}, which are closely
related to RM codes. For a performance comparison between polar and
RM codes, see \cite{Ari10, arikan:rmpolarcomp}.  Simulations
and analytical results suggest that RM codes do not perform well
under successive and iterative decoding, but they
outperform  polar  codes  under maximum a posteriori (MAP) decoding
\cite{Ari09, hussami:perfo}. 
%In addition, an interpolation method
%between polar and RM codes was proposed in \cite{MoHaUr14} in order
%to improve the performance of polar codes not only under MAP decoding,
%but also under low-complexity decoding schemes. 
Nevertheless, it
is not known whether RM codes themselves are capacity-achieving
except for rates approaching $0$~and~$1$ over the binary erasure
channel (BEC) and the binary symmetric channel (BSC)~\cite{abbe}.

In this paper, we show that RM codes indeed achieve the capacity for
transmission over the BEC for \emph{any} rate $R\in (0, 1)$.  The same
result was shown independently by Kumar and Pfister \cite{KuP15} using
essentially the same approach.

%----------------------------------------------------
\section{Main Result} \label{sec:main}
%----------------------------------------------------
Let RM$(n, r)$ denote the Reed--Muller (RM) code of {\em block
length} $N=2^n$ and {\em order} $r$, see \cite{sloane}.  This is a
linear code of rate $R=\frac1N \sum_{i=0}^r \binom{n}{i}$ and minimum
distance $d=2^{n-r}$, generated by all rows of weight at least
$2^{n-r}$ of the Hadamard matrix $\left(\begin{smallmatrix}1&0\\
1&1\end{smallmatrix}\right)^{\otimes n}$, where $\otimes$ denotes
the Kronecker product. Let $[N]=\{1,\dotsc,N\}$ denote the index
set of codeword bits.  For $i\in[N]$, let $x_i$ denote
the $i$th component of a vector $x$, and let $x_{\sim i}$ denote
the vector containing all components \emph{except} $x_i$.  For $x,
y \in \{0, 1\}^N$, we write $x \prec y$ if $y$ dominates $x$
component-wise, i.e. if $x_i \le y_i$ for all $i\in [N]$.

Let BEC$(\epsilon)$ denote the binary erasure channel with
erasure probability $\epsilon$.  Recall that this channel has {\em
capacity} $1-\epsilon$ bits/channel use.  In what follows, we will fix
a rate $R$ for a sequence of RM codes and show that the bit error
probability of the code sequence vanishes for all BECs with capacity
strictly larger than~$R$, i.e., erasure probability strictly smaller
than~$1-R$.

\begin{theorem}[RM Codes Achieve Capacity on the BEC]\label{the:rmbec}
Consider a sequence of RM$(n, r_n)$ codes of increasing $n$ and
rate $R_n$ converging to $R$, $0 < R < 1$.  For any $0 \leq
\epsilon < 1-R$ and any $\delta>0$ there exists an $n_0$ such that
for all $n>n_0$ the \emph{bit error probability} of 
RM$(n, r_n)$ is bounded above by $\delta$ under \emph{bit-MAP} decoding.  
\end{theorem}

The only property of RM codes that has a bearing on the following
proof of Theorem~\ref{the:rmbec} is that these codes exhibit a high
degree of symmetry, and in particular, that they are invariant under a
2-transitive group of permutations on the coordinates of the code
\cite{KLP68:2transRM,sloane,Berger93}.  In fact, this proof also shows
that all 2-transitive sequences of codes are capacity-achieving. We
will return to this point in Section~\ref{sec:general}. 

\begin{lemma}[RM Codes Are $2$-Transitive]\label{lem:2transitivity}
For any $a$, $b$, $c$, and $d\in[N]$ s.t. $a\neq b$ and $c\neq d$, there exists a
permutation $\pi:[N]\to[N]$ such that 
\begin{itemize}
\item[(i)]
$\pi(a)=c$, $\pi(b)=d$, and
\item[(ii)]
$RM(n,r)$ is closed under the permutation of its codeword bits according to
$\pi$. That is, 
\begin{equation}
\begin{split}
(x_1,\dotsc,x_N)&\in RM(n,r)\\
&\Updownarrow\\
(x_{\pi(1)},\dotsc,x_{\pi(N)})&\in RM(n,r).
\end{split}
\end{equation}
\end{itemize}
\end{lemma}

%I'M COMMENTING OUT THE FOLLOWING PARAGRAPH BECAUSE THE STATEMENT IS
%NOT EXACTLY CORRECT AND THERE'S NO EASY WAY TO FIX IT WITHOUT TOO
%MANY WORDS OR ESSENTIALLY REPEATING THE ABOVE LEMMA, WHICH ITSELF IS
%CLEAR ENOUGH [EREN]

%A subset of $\{0, 1\}^N$ is said to be 2-transitive if it is closed
%under any permutation of its positions such that two values of the
%permutation are assigned. In the case of the previous lemma, the
%subset of $\{0, 1\}^N$ consists of the codewords of a RM code, and the
%two assigned values of the permutation are $\pi(a)=c$, $\pi(b)=d$.
%Similarly, a subset of $\{0, 1\}^N$ is said to be 1-transitive, if in
%the previous lemma (i) and (ii) hold, but where in (i) we only require
%a single condition (e.g., $\pi(a)=c$).

The $2$-transitivity of the code implies many symmetries that will
be critical in the proof, which we outline here. We will be interested
in MAP decoding of the $i$th codebit $x_i$ from observations
$y_{\sim i}$, that is, all channel outputs except $y_i$.  The error
probability of the $i$th such decoder for transmission over a
BEC$(\epsilon)$ is called the $i$th \emph{EXIT function}~\cite[Lemma
3.74]{RiU08}, which we denote by $h_i(\epsilon)$. We will see that
all $N$ EXIT functions of an RM code (and of any $2$-transitive
code) are identical, and also that erasure patterns that lead to
decoding errors under this decoder exhibit a high degree of symmetry.
These symmetries will imply that the EXIT functions have a sharp
threshold behavior, i.e., the bit error probability is very small
below a threshold, and very large above.  A final and crucial benefit
of considering this suboptimal decoder and EXIT functions instead
of the optimal block-MAP decoder is the well-known \emph{Area
Theorem}~\cite{Ash02,Ash04,Mea04, RiU08}, which will allow us to
show that the threshold is at channel capacity and conclude the
proof.

Recall the basic definition of an EXIT function \cite[Lemma
3.74]{RiU08} and its relation to bit-MAP decoding.

\begin{definition}[EXIT Function]\label{def:exit} Let $\code[N, K]$
be a binary linear code of rate $R=K/N$ and let $X$ be chosen with
uniform probability from $\code[N, K]$.  Let $Y$ denote the result
of letting $X$ be transmitted over a BEC$(\epsilon)$.  The EXIT
function $h_i(\epsilon)$ associated with the $i$th bit of $\code$
is defined as \begin{equation} h_i(\epsilon) =
H(X_i \mid Y_{\sim i}).  \end{equation} \end{definition}

\begin{lemma}[EXIT Function and Bit-MAP Decoding]\label{lem:exit}
Let $\code[N, K]$
be a binary linear code
and let $\hat{x}^{\text{\tiny
MAP}}(y_{\sim i})$ denote the MAP estimator of the $i$th code
bit given the observation $y_{\sim i}$. Then, \begin{equation}
h_i(\epsilon) = \prob(\hat{x}^{\text{\tiny MAP}}(Y_{\sim i}) = ?).
\end{equation} 
\end{lemma}

The most relevant property of EXIT functions for our purpose is the Area Theorem,
see \cite{Ash02,Ash04,Mea04, RiU08}.
\begin{lemma}[Area Theorem]\label{lem:areatheorem}
Let $\code[N, K]$
be a binary linear code, and let $h(\epsilon)=\frac1N
\sum_{i=0}^{N-1} h_i(\epsilon)$ be the {\em average} EXIT function.
Then,
\begin{align*}
\int_{0}^{\epsilon} h(x) \;\text{d}x = \frac1N H(X \mid Y),
\end{align*}
where $H(X \mid Y)$ is the conditional entropy of the codeword $X$
given the observation $Y$ at the receiver. In particular,
\begin{align*}
\int_{0}^{1} h(x) \;\text{d}x = R=\frac{K}{N}.
\end{align*}
\end{lemma}

We now show that the erasure patterns that lead to decoding failures
are monotone and symmetric.  Recall that the decoding of each bit
relies only on $N-1$ received bits.  We will denote each erasure
pattern by a binary vector of length $N-1$, where a~$1$ denotes an
erasure and a~$0$ denotes a non-erasure. We first characterize the set
$\Omega_i$ that leads to a decoding failure for bit~$i$.

\begin{definition}[$\Omega_i$]\label{def:omegai} 
Given a binary linear code $\code[N, K]$, let $\Omega_i$ be the set that consists of all
$\omega\in\{0,1\}^{N-1}$ for which there exists $c\in\code$ such that
$c_i=1$ and $c_{\sim i}\prec\omega$.
\end{definition}

\begin{lemma}[$\Omega_i$ Encodes $h_i(\epsilon)$]\label{lem:encoding}
Let $\omega\in\{0,1\}^{N-1}$ be the erasure pattern on the received
bits $y_{\sim i}$.  Then the $i$th bit-MAP decoder fails if and only if
$\omega\in\Omega_i$.  Consequently, if $\mu_{\epsilon}(\cdot)$ is the
measure on $\{0, 1\}^{N-1}$ that puts weight $\epsilon^w
(1-\epsilon)^{N-1-w}$ on a point of Hamming weight $w$, then
\begin{align*}
h_i(\epsilon)=\mu_{\epsilon}(\Omega_i).
\end{align*}
That is, $\Omega_i$ ``encodes'' the EXIT function of the $i$th position. 
\end{lemma}

\begin{IEEEproof}
%We first note that a codeword $\cw$ has $N$ bits but $\omega$ is only of
%length $N-1$. This is because when we compute the EXIT function we
%disregard the position $i$ itself and are only given the partial
%observation $Y_{\sim i}$. So the received value at position $i$ is
%irrelevant.
%
%We claim that $\Omega_i$ is in one-to-one correspondence with the
%set of erasure patterns so that the codebit $i$ cannot be
%decoded by a MAP decoder that is given the partial observation $Y_{\sim
%i}$. More precisely, every element $\omega$ of $\Omega_i$ is a point
%in $\{0, 1\}^{N-1}$ and we think of this $\omega$ as the characteristic
%vector that encodes the erasure set -- there is an erasure wherever
%$\omega$ has a $1$.
Since the code is linear and the channel is symmetric and memoryless, we can assume that the all-zero codeword was transmitted. Given an erasure pattern $\omega$, let $\code'$ denote the set of all codewords $c$ that are compatible with the observation $y_{\sim i}$,
i.e., all codewords for which $c_{\sim i}\prec\omega$. Note that
since the code is linear, so is $\code'$.  This implies that if there
exists a $c\in\code'$ with $c_i=1$, then half of all codewords in
$\code'$ have a~$0$ at position $i$, and the other half have a~$1$,
and thus the bit-MAP decoder fails to decode bit $i$. On the other hand, if
there is no $c\in\code'$ with $c_i=1$, then all compatible codewords
have a $0$ at position $i$, and thus the bit-MAP decoder succeeds.
That is, $\Omega_i$ is the set of all erasure patterns s.t. the bit-MAP decoder cannot decide on position $i$ given the observation $y_{\sim i}$. The claim that
$h_i(\epsilon)=\mu_{\epsilon}(\Omega_i)$ follows immediately,
since the memorylessness of the channel implies that an erasure
pattern $\omega$ occurs with probability $\mu_\epsilon(\omega)$. 
\end{IEEEproof}

\begin{lemma}[$\Omega_i$ is Monotone]\label{lem:monotonicity}
If $\omega \in \Omega_i$ and $\omega \prec
\omega'$, then $\omega' \in \Omega_i$.  \end{lemma}
\begin{IEEEproof}
If $\omega \in \Omega_i$, then there exists a
codeword $\cw$ so that $\cw_i=1$ and $\cw_{\sim i} \prec
\omega$.  Since by assumption $\omega \prec \omega'$, it follows
that $\cw_{\sim i} \prec \omega'$, which implies $\omega' \in \Omega_i$.  
\end{IEEEproof}

\begin{lemma}[$\Omega_i$ is Symmetric]\label{lem:invariance} 
If $\code[N,K]$ is a $2$-transitive binary linear code,
then $\Omega_i$ is invariant under a $1$-transitive
group of permutations for any $i\in [N]$. Following \cite{Friedgut96}, we say that $\Omega_i$ is {\em
symmetric}.
\end{lemma} 
\begin{IEEEproof} 
Since $\code$ is $2$-transitive, for any $j_1, j_2 \in[N]\setminus
\{i\}$, there exists a permutation $\pi: [N]\to [N]$ so that
\begin{itemize}
\item $\pi(i)=i$,
\item $\pi(j_1)=j_2$,
\item $(c_{\pi(1)},\dotsc,c_{\pi(N)})\in \code$ for any $(c_1,\dotsc,c_N)\in \code$.
\end{itemize}
Let $S_1:[N-1]\to [N]\setminus \{i\}$ be defined as $S_1(k)=k$ for $k\in \{1, \cdots, i-1\}$ and $S_1(k)=k+1$ for $k\in \{i, \cdots, N-1\}$. Let $S_2: [N]\setminus \{i\} \to [N-1]$ be defined as $S_2(k)=k$ for $k\in \{1, \cdots, i-1\}$ and $S_2(k)=k-1$ for $k\in \{i+1, \cdots, N\}$. Consider the permutation $\hat{\pi}: [N-1]\to [N-1]$ defined as $\hat{\pi}(k)= S_2(\pi(S_1(k)))$. Note that, by changing the choice of $j_1$ and $j_2$, we generate the $1$-transitive group of permutations on $[N-1]$. It then suffices to show that if $\omega = (\omega_1, \cdots, \omega_{N-1}) \in \Omega_i$,
then $(\omega_{\hat{\pi}(1)}, \cdots, \omega_{\hat{\pi}(N-1)}) \in \Omega_i$.

Recall that $\omega \in \Omega_i$ if there exists a codeword $\cw=(c_1,\dotsc,c_N)\in \code$
so that $\cw_i=1$ and $\cw_{\sim i} \prec \omega$. By construction of $\pi$, we have that $(c_{\pi(1)},\dotsc,c_{\pi(N)})\in \code$ and, in addition,
$\cw_{\pi(i)}=\cw_i=1$. By construction of $\hat{\pi}$, $(c_{\pi(1)}, \cdots, c_{\pi(i-1)}, c_{\pi(i+1)}, \cdots, c_{\pi(N)}) \prec
(\omega_{\hat{\pi}(1)}, \cdots, \omega_{\hat{\pi}(N-1)})$. As a result, $(\omega_{\hat{\pi}(1)}, \cdots, \omega_{\hat{\pi}(N-1)}) \in \Omega_i$ and the proof is complete.
\end{IEEEproof}

We now show that all EXIT functions of a $2$-transitive code are 
identical.
\begin{lemma}[$h_i$ is Independent of $i$]\label{lem:independence}
If $\code[N,K]$ is a $2$-transitive binary linear code, then
$h_i(\epsilon)=h_j(\epsilon)$ for all $i, j \in[N]$.  That is, $h_i(\epsilon)$ is independent of $i$.  
\end{lemma} 
\begin{IEEEproof} 
Since $\code$ is 2-transitive, there exists a permutation $\pi:[N]\to[N]$ so that 
\begin{itemize}
\item $\pi(i)=j$,
\item $(c_{\pi(1)},\dotsc,c_{\pi(N)})\in \code$ for any $(c_1,\dotsc,c_N)\in \code$.
\end{itemize} 
Let $S_i: [N-1]\to [N]\setminus \{i\}$ be defined as $S_i(k)=k$ for $k\in \{1, \cdots, i-1\}$ and $S_i(k)=k+1$ for $k\in \{i, \cdots, N-1\}$. Let $S_j:[N]\setminus \{j\}\to [N-1]$ be defined as $S_j(k)=k$ for $k\in \{1, \cdots, j-1\}$ and $S_j(k)=k-1$ for $k\in \{j+1, \cdots, N\}$. Consider the permutation $\hat{\pi}: [N-1]\to [N-1]$ defined as $\hat{\pi}(k)= S_j(\pi(S_i(k)))$.

Pick $\omega \in \Omega_j$. Then, there exists a codeword $\cw$ so that $\cw_j = 1$ and $\cw_{\sim
j} \prec \omega$. By construction of $\pi$, we have that $(c_{\pi(1)},\dotsc,c_{\pi(N)})\in \code$ and, in addition,
$\cw_{\pi(i)}=\cw_j=1$. By construction of $\hat{\pi}$, $(c_{\pi(1)}, \cdots, c_{\pi(i-1)}, c_{\pi(i+1)}, \cdots, c_{\pi(N)}) \prec
(\omega_{\hat{\pi}(1)}, \cdots, \omega_{\hat{\pi}(N-1)})$. As a result, $(\omega_{\hat{\pi}(1)}, \cdots, \omega_{\hat{\pi}(N-1)}) \in \Omega_i$. 

With an abuse of notation, let us define 
$$\hat{\pi}(\Omega_j)=\{(\omega_{\hat{\pi}(1)}, \cdots, \omega_{\hat{\pi}(N-1)}) : \omega \in \Omega_j\}.$$
Then, the previous argument implies that $\hat{\pi}(\Omega_j)\subseteq \Omega_i$.

It is clear that, if $\omega\neq \omega'$, then $(\omega_{\hat{\pi}(1)}, \cdots, \omega_{\hat{\pi}(N-1)})\neq (\omega'_{\hat{\pi}(1)}, \cdots, \omega'_{\hat{\pi}(N-1)})$. Indeed, if $\omega\neq \omega'$, then there exists an index $k$ s.t. $\omega_{k}\neq \omega'_{k}$ and, therefore, $\omega_{\hat{\pi}(k)}\neq \omega'_{\hat{\pi}(k)}$. In addition, the permutation $\hat{\pi}$ leaves the weight of $\omega$ unchanged. As a result, we have
\begin{equation}
h_j(\epsilon) = \mu_{\epsilon}(\Omega_j)\stackrel{\mathclap{\mbox{\footnotesize(a)}}}{=}\mu_{\epsilon}(\hat{\pi}(\Omega_j)) \stackrel{\mathclap{\mbox{\footnotesize(b)}}}{\le} \mu_{\epsilon}(\Omega_i) = h_i(\epsilon),
\end{equation}
where (a) comes from the fact that the channel acts independently and identically on each component, and (b) follows from $\hat{\pi}(\Omega_j)\subseteq \Omega_i$. By repeating the same argument with the indices $i$ and $j$ exchanged, we obtain opposite inequality and, therefore, the thesis follows.    
\end{IEEEproof}

We recall here the main ingredient for our proof, due to Friedgut
and Kalai. We note that Tillich and Z\'{e}mor applied the following theorem
in \cite{TiZ00} to show that {\em every} sequence of linear codes
of increasing Hamming distance has a sharp threshold under block-MAP
decoding for transmission over the BEC and the BSC.

\begin{theorem}[Sharp Threshold -- \cite{Friedgut96}]\label{the:sharpthreshold}
Let $\Omega\in\{0,1\}^N$ be a symmetric monotone set, where symmetry
and monotonicity are defined as in 
Lemma
\ref{lem:monotonicity}~and~\ref{lem:invariance}, respectively. 
If $\mu_{\underline{\epsilon}}(\Omega) > \delta$,
then $\mu_{\overline{\epsilon}}(\Omega) > 1-\delta$ for
$\overline{\epsilon} = \underline{\epsilon} + c \frac{\log(\frac{1}{2
\delta})}{\log(N)}$, where $c$ is an absolute constant.  \end{theorem}

\begin{IEEEproof}[Proof of Theorem~\ref{the:rmbec}]
Consider a sequence of codes RM$(n, r_n)$ with rates converging to
$R$.  That is, the $n$th code in the sequence has a rate $R_n\le
R+\delta_n$, where $\delta_n\to0$ as $n\to \infty$. 

Lemma~\ref{lem:independence} implies that $h_i(\epsilon)$ is
independent of~$i$, and, thus, it is equal to the average EXIT function
$h(\epsilon)$. Therefore, by Lemma~\ref{lem:areatheorem} we have
$$\int_{0}^{1} h_i(\epsilon) \;\text{d}\epsilon=R_n \leq R +
\delta_{n}.$$

Consider the set
$\Omega_i$ defined in Definition \ref{def:omegai} that encodes $h_i(\epsilon)$. By Lemmas
\ref{lem:monotonicity}~and~\ref{lem:invariance}, $\Omega_i$ is monotone and symmetric. Therefore, from Lemma~\ref{the:sharpthreshold} we have that if
$h_i(\overline{\epsilon}) = 1-\delta$,  then
$h_i(\underline{\epsilon})\le\delta$ for $\overline{\epsilon} =
\underline{\epsilon} + c \displaystyle\frac{\log(\frac{1}{2 \delta})}{\log(N-1)}$, where $c$ is an absolute constant.

Now, the function $h_i(\epsilon)$ is increasing, and therefore by
Lemma 2, the error probability of the $i$th bit-MAP decoder is upper
bounded by $\delta$ for all $i\in [N]$ and $\epsilon \le \underline{\epsilon}$. In order to conclude the proof, it suffices to show that $\underline{\epsilon}$ is close to $1-R$.
Note that by definition of $\overline{\epsilon}$, the area under
$h_i(\epsilon)$ is at least equal to
\begin{align*}
(1-\overline{\epsilon})(1-\delta) \geq 1-\overline{\epsilon}-\delta =
1- \underline{\epsilon} - c \frac{\log(\frac{1}{2
\delta})}{\log(N-1)} -\delta.
\end{align*}
On the other hand, this area is at most equal to
$R+\delta_{n}$.  Combining these two inequalities we obtain
\begin{align} \label{equ:epsilonbound}
\underline{\epsilon} \geq 1-R -\delta-\delta_{n}  - c
\frac{\log(\frac{1}{2 \delta})}{\log(N-1)}.
\end{align}
We see that $\underline{\epsilon}$ can be made arbitrarily close to
$1-R$ by picking~$\delta$ sufficiently small and~$N$ sufficiently
large. That is, the bit error probability can be made arbitrarily
small at rates arbitrarily close to $1-R$.  
\end{IEEEproof}

%----------------------------------------------------
\section{Generalizations and Discussion} \label{sec:general}
%----------------------------------------------------

As mentioned above, the foregoing arguments hold for all $2$-transitive
codes, and not just RM codes.  That is, all such codes are capacity
achieving over the BEC under bit-MAP decoding.  This includes, for
example, the class of extended BCH codes (\cite[Chapter 8.5, Theorem
16]{sloane}).

RM codes are only one possible family of codes that can be derived from
the Hadamard matrix. It is reasonable to assume that any subset of
generators of sufficient weight from the Hadamard matrix will produce
good codes.  It would be interesting to see if such a statement can be
proved.  Clearly, the symmetries of RM codes that are used here will
not be present in general. 

Perhaps of even greater interest is whether RM codes achieve capacity
on general binary-input memoryless output-symmetric channels and if
the above technique can be extended.  Note that it suffices to prove
that RM codes achieve capacity {\em for the BSC} since (up to a small
factor) the BSC is the worst channel, see \cite[pp.\
87--89]{squirrel:thesis}.   Most of the notions that we used here for
the BEC have a straighforward generalization (e.g., GEXIT functions
replace EXIT functions) or need no generalization ($2$-transitivity).
However, it is currently unclear if the GEXIT function can be encoded
in terms of a monotone function. It is likely that different
techniques will be needed to show sharp thresholds for GEXIT
functions.

One of the main motivations for studying RM codes is their superior
empirical performance (over the BEC) compared with the
capacity-achieving polar codes.  By far the most important practical
question is whether this promised performance can be harnessed at low
complexities.

%----------------------------------------------------
\section*{Acknowledgement}
%----------------------------------------------------
This work was done while the authors were visiting the Simons
Institute for the Theory of Computing, UC Berkeley. We would like to
thank the Institute for providing us with a fruitful work environment.
We further gratefully acknowledge discussions with Tom Richardson,
Hamed Hassani, and in particular with Henry Pfister.

\bibliographystyle{IEEEtran}
\bibliography{biblio}

\end{document}